\documentclass[journal]{IEEEtran}
\ifCLASSINFOpdf
\else
\usepackage[dvips]{graphicx}
\graphicspath{{Figs/}}
\DeclareGraphicsExtensions{.eps}
\fi
\usepackage[cmex10]{amsmath}
\begin{document}
%
\title{Plasma walls beyond the perfect absorber approximation for electrons}
%
%
%

\author{Franz~X. Bronold,
        Rafael~L. Heinisch,
        Johannes~Marbach,
        and Holger~Fehske  
\thanks{The authors are with the Institute of Physics, 
Ernst-Moritz-Arndt-University Greifswald, D-17489 Greifswald, 
Germany, e-mail: bronold@physik.uni-greifswald.de.}
\thanks{Manuscript received September 15, 2010; revised xx, xx.}}

%
%

\markboth{Journal of \LaTeX\ Class Files,~Vol.~6, No.~1, January~2007}%
{Shell \MakeLowercase{\textit{et al.}}: Bare Demo of IEEEtran.cls for Journals}
%



\maketitle

\begin{abstract}
Plasma walls accumulate electrons more efficiently than ions leading to wall potentials 
which are negative with respect to the plasma potential. Theoretically, walls are usually 
treated as perfect absorber for electrons and ions implying perfect sticking of the particles to 
the wall and infinitely long desorption times for particles stuck to the wall. For electrons 
we question the perfect absorber model and calculate, specifically for 
a planar dielectric wall, the electron sticking coefficient $s_e$ and the electron 
desorption time $\tau_e$. For the uncharged wall we find $s_e\ll 1$ and $\tau_e\approx 10^{-4}s$. 
Thus, in the early stage of the build-up of the wall potential, when the wall is essentially 
uncharged, the wall is not a perfect absorber for electrons. For the charged wall we find 
$\tau_e^{-1}\approx 0$. Thus, $\tau_e$ approaches the perfect absorber value. But $s_e$
is still only of the order of $10^{-1}$. Calculating $s_e$ as a function of the wall
potential and combining this expression with the quasi-stationary balance equations 
for the electron and ion surface densities we find the selfconsistent wall potential, 
including surface effects, to be $30\%$ of the perfect absorber value.
\end{abstract}

\begin{IEEEkeywords}
Plasma-sheath, plasma-wall interaction, wall-charging 
\end{IEEEkeywords}

%
\IEEEpeerreviewmaketitle

\section{Introduction}
%
%
%
%
\IEEEPARstart{M}{acroscopic} objects in contact with an ionized gas 
-- plasma walls -- act as sinks and sources for the charged and 
uncharged particles of the plasma. Because electrons are collected more 
efficiently than ions, walls are negatively charged and thus shielded
from the bulk plasma by a space charge depletion layer (plasma sheath).
But not only the spatial homogeneity of the plasma is strongly affected 
by the wall. Surface-supported electron-ion recombination and secondary
electron emission severely modify the overall charge balance of the discharge. 
Particularly in dusty plasmas~\cite{Ishihara07,FIK05,Mendis02} and  
solid-state-based microdischarges~\cite{Kushner05,BSE06} the wall becomes 
an integral part of the plasma. 

The microscopic understanding of the build-up of the negative wall potential
is in a rather rudimentary stage. It is usually based on the assumption that
electrons and ions hitting the wall are instantaneously annihilated which is 
the same as to say the wall is a perfect absorber for electrons and ions. 
The wall potential arising from this picture is the one which equalizes at 
the wall the electron and ion in-fluxes from the plasma~\cite{Franklin76}. 
The presence of surface charges which is clearly necessary for a wall 
potential to develop is hard to reconcile with the instantaneous annihilation
assumption. It is moreover almost always assumed that secondary electrons
from the wall are released from the electronic bulk states and not from the 
electronic surface states which should in fact host the electrons accumulated 
from the plasma.

Following the lead of Emeleus and Coulter~\cite{EC87,EC88} and 
others~\cite{BBD97,USB00,GMB02} who, respectively, introduced and applied the 
idea of a two-dimensional electron surface plasma attached to plasma walls, we 
recently proposed to visualize the charging of plasma walls as an electron 
physisorption process~\cite{BFKD08,BDF09}. In the surface-plasma based physisorption 
scenario the wall potential arises from two-dimensional electron and ion surface 
densities which, for a collisionless planar sheath, obey two coupled balance 
equations, 
\begin{align}
\frac{dn_e}{dt}=&s_ej^{\rm th}_e-\frac{n_e}{\tau_e}-\alpha_{\rm rw}n_in_e~,
\label{ne}\\
\frac{dn_i}{dt}=&s_ij_i^B-\frac{n_i}{\tau_i}-\alpha_{\rm rw}n_in_e~,
\label{ni}
\end{align}
where $j^{\rm th}_e$ and $j_i^B$ are, respectively, the thermal electron and the mono-energetic
ion in-flux from the plasma. The surface properties are thereby encoded in the electron and ion 
sticking coefficients $s_{e,i}$, the electron and ion desorption times $\tau_{e,i}$, and 
the wall recombination constant $\alpha_{\rm rw}$. At quasi-stationarity, Eqs. (\ref{ne}) 
and (\ref{ni}) reduce to 
\begin{align}
s_ej^{\rm th}_e=s_ij^B_i+\frac{n_e}{\tau_e}-\frac{n_i}{\tau_i}~,
\label{balance}
\end{align}
which is a selfconsistency equation for the wall potential $\phi_w$, which enters
through the thermal electron flux $j^{\rm th}_e$ and the electron surface density 
$n_e$. The perfect absorber approximation corresponds to $s_{e,i}=1$ and 
$\tau_{e,i}^{-1}=0$. To improve this approximation one either has to 
measure $s_{e,i}$ and $\tau_{e,i}$ directly or calculate these quantities from 
microscopic models for the electron(ion)-wall interaction. Both is challenging. 
But advanced non-invasive techniques of measuring surface charges~\cite{KSC02,SLP07} 
may successfully guide the construction of realistic microscopic models for the plasma 
wall.

We expect $s_e$ and $\tau_e$ to be particularly important parameters, especially in the 
early stages of the build-up of the wall potential. Using simple quantum-mechanical models 
for the electron-surface interaction we calculated therefore these two quantities for 
uncharged metallic~\cite{BDF09} and uncharged dielectric~\cite{HBF10a,HBF10b} surfaces 
and found surprisingly small electron sticking coefficients. Only for metallic surfaces 
was the product $s_e\tau_e$ in the range expected from studies of dc column 
plasmas~\cite{BBD97,USB00} and grain charging~\cite{BFKD08}. 

Since we calculated $s_e$ and $\tau_e$ only for uncharged surfaces our previous results 
are only applicable to the very beginning of the charging 
process, when the wall is basically uncharged. Below we will extend our microscopic 
considerations to charged plasma walls. The electronic states on which the calculation 
of $s_e$ and $\tau_e$ has to be based are then no longer the polarization-induced 
external surface states (image states) we employed for uncharged surfaces but 
unoccupied internal conduction band states of the wall. Nevertheless, the build-up
of the wall potential can still be considered as an electron physisorption process. 

In the next section we qualitatively discuss general microscopic aspects of the electron-wall 
interaction. To be specific we restrict ourselves to a planar dielectric wall. We then
recall briefly in Section III the theoretical approach we employed previously to 
calculate $s_e$ and $\tau_e$ for uncharged surfaces and present representative 
results for graphite and MgO. In Section IV we describe a strategy to estimate 
$s_e$ and $\tau_e$ for charged dielectric walls. Numerical results are given for 
a sapphire wall (${\rm Al_2O_3}$). We then combine our expressions for 
$\tau_e^{-1}$ and $s_e$ with Eq.~(\ref{balance}), setting $s_i=1$ and $\tau_i^{-1}=0$, 
to calculate the selfconsistent wall potential beyond the perfect absorber model for
electrons. It turns out to be roughly one-third of the perfect absorber value.
Finally, Section V gives the conclusions we draw from our results.

\section{Electron-Wall interaction}

To discuss the microscopic aspects of the electron-wall interaction we consider a planar 
dielectric wall. It defines the $xy$-plane of a coordinate system separating the solid in 
the halfspace $z\le 0$ from the plasma in the halfspace $z>0$. 

Quite generally, a quantum-mechanical calculation of the electron sticking coefficient 
$s_e$ and electron desorption time $\tau_e$ has to be based on a Hamiltonian,
\begin{eqnarray}
H=H_{e}+H_{w}+H_{e-w}~,
\label{Hmodel}
\end{eqnarray}
where $H_{e}$, $H_w$, and $H_{e-w}$ describe, respectively, the unperturbed dynamics of an electron 
in the vicinity of the wall, the elementary excitations of the wall responsible for electron energy 
relaxation, and the coupling between the two. 

The electronic structure in the vicinity of the wall is rather complex. It depends on 
the plasma and the surface. 
Assuming, for simplicity, a 
perfect boundary, $H_w$ is a single-electron Hamiltonian belonging to the electron potential 
energy
\begin{align}
V(z)=\left\{\begin{array}{ll}
V_{\rm c}(z) & ~~{\rm for}~~ z\le 0\\
V_{\rm p}(z)+V_{\rm s}(z) & ~~{\rm for}~~ z>0~,
\end{array}\right.
\label{V(x)}
\end{align}
where $V_{\rm c}$ is the crystal potential of the wall material, $V_{\rm p}$ is the 
exchange- and correlation-induced polarization potential which confines the electrons inside
the material and causes the attraction of external electrons to the surface at short distances, 
and $V_{\rm s}$ is the potential energy in the sheath which leads to a Coulomb barrier 
for electrons approaching the wall from the plasma. As explained in Ref.~\cite{BDF09} $V(z)$ 
supports volume states (periodic inside the wall and exponentially decaying in the plasma), bound 
and unbound surface states (the former exponentially decaying on both sides of the plasma-wall
interface and the latter decaying only inside the wall), as well as free states (non-decaying
on both sides of the interface). 

\begin{figure}[t]
\centering
\includegraphics[width=0.98\linewidth]{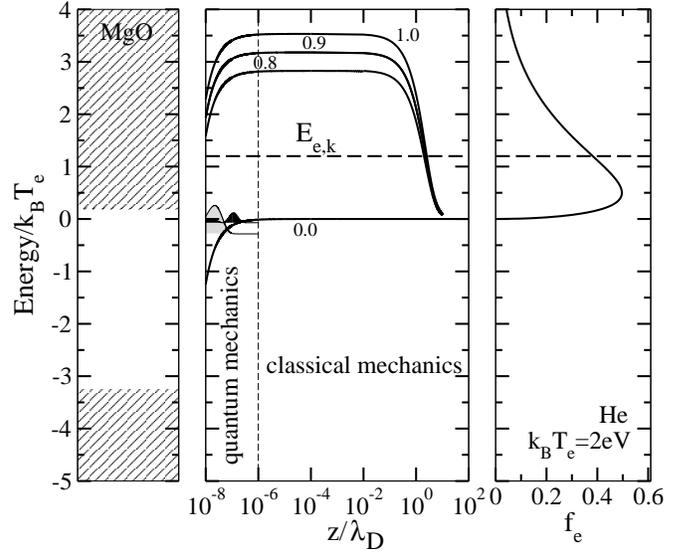}
\caption{The middle panel shows the potential energy of an electron
in a collisionless sheath in front of a negatively charged dielectric 
boundary. Close to the boundary the potential energy decreases because of 
the polarization-induced attraction. The number attached to the 
graphs give the wall potential in units of the perfect absorber 
value, Eq. (\ref{phiw}). The material parameters are appropriate for 
a MgO wall, whose band structure is schematically shown in the left panel, 
and a He discharge with $N_0=10^7cm^{-3}$ and $k_BT_e=2eV$. The electron 
energy distribution function $f_e(E)$ in the bulk of the discharge 
is plotted on the right side of the panel. In the main panel are also 
shown the two lowest image states controlling physisorption of an 
electron at an uncharged MgO surface. The processes close to or in the 
wall have to be described quantum-mechanically whereas the physics 
of the discharge is of course classical.}
\label{Sheath}
\end{figure}

A plasma electron approaching the wall may get trapped (adsorbed) if it can get rid 
of its excess energy via inelastic scattering processes. Once it is trapped it may 
de-trap again (desorb) if it gains enough energy from the wall. The scattering processes 
depend on the wall material. For dielectric walls, which have large energy gaps, optical 
and acoustic phonons cause energy relaxation whereas electron-hole pairs and plasmons
trigger energy relaxation at metallic walls. The matrix elements of the coupling depend 
on whether surface or volume states are involved in the scattering process and thus 
on the microscopic details of the interface and the number of electrons already 
collected by the wall.  

To determine what kind of electronic states are most likely involved in the build-up
of the wall potential it is instructive to consider the potential energy on the plasma 
side of the plasma-wall interface. For a collisionless sheath, it is given by
\begin{align}
V(z>0)=
k_BT_e\eta_s(z)-\frac{1}{4}\frac{\epsilon_s-1}{\epsilon_s+1}\frac{e^2}{z}
\label{potentialE}
\end{align}
where $\epsilon_s$ is the dielectric constant of the wall, $e$ is the elementary charge, and
$\eta_s(z)$ is the solution of~\cite{Franklin76}
\begin{align}
\lambda_D^2\frac{d^2\eta_s}{dz^2}=\frac{1}{\sqrt{1+\frac{2\eta_s}{u_{i0}^2}}}-\exp[-\eta_s]~,
\label{sheathDGL}
\end{align}
with $\lambda_D^2=k_BT_e/eN_0$ the Debye screening length, $u_{i0}=v_i/\sqrt{k_BT_e/m_i}$ 
the ion velocity, and $N_0$ the volume plasma density; $m_i$ is the ion mass and 
$k_BT_e$ is the mean electron energy in the plasma.

Figure \ref{Sheath} shows $V(z>0)$ for a MgO surface ($\epsilon_s=9.8$) in contact 
with a He discharge whose mean electron energy $k_BT_e=2eV$ and bulk plasma density is 
$N_0=10^{-7}cm^{-3}$. To solve Eq.~(\ref{sheathDGL}) we used the perfect absorber boundary condition, 
that is, we calculated the wall potential from $j_e^{\rm th}=j_i^B$ which leads to
\begin{align}
-e\phi_w=\frac{1}{2}k_BT_e\ln\big[\frac{m_i}{2\pi m_e}\big]~.
\label{phiw}
\end{align}
To mimic the build-up of the wall potential we multiplied the perfect absorber value
by the numerical factor attached to the graphs. 

As can be seen close to the boundary the potential energy decreases because of the polarization-induced 
attractive short-range part of the potential energy, the second term on the right hand side of 
Eq.~(\ref{potentialE}). Since MgO has a negative electron affinity 
$-\chi=0.2eV-0.4eV$~\cite{BKP07,RWK03} the vacuum level is below 
the conduction band edge. Image states are thus possible and should control electron physisorption at 
the uncharged MgO surface. Once the surface is charged the Coulomb barrier due to the sheath 
potential allows however only electrons with kinetic energy larger than the Coulomb repulsion to 
reach the wall. In that energy range image states are unstable and the volume states indicated in 
the left panel of Fig. \ref{Sheath} are expected to be most important for adsorption to and 
desorption from the wall.

Having identified the physical processes leading, on the microscopic scale, to the build-up of a wall 
potential, we can now attempt a quantum mechanical calculation of the electron desorption time $\tau_e$ 
and the electron sticking coefficient $s_e$. This will be the topic of the next two sections.

\section{Uncharged dielectric surfaces}

If the electron affinity of an uncharged dielectric surface is negative, electron trapping 
and de-trapping occurs in polarization-induced external surface states (image states). In 
a theoretical approach patterned on that of physisorption of atoms and molecules~\cite{KG86}
we calculated in Refs.~\cite{HBF10a,HBF10b} $s_e$ and $\tau_e$ for such a situation. For 
completeness we recall in this section the main features of our approach and discuss 
representative data. 

The starting point is a quantum-kinetic equation for the occupancies of the image 
states (Eq. (3) in Ref.~\cite{HBF10b}), 
\begin{align}
\frac{\mathrm{d}}{\mathrm{d}t}n_n(t)=&\sum_{n^\prime} 
     \left[{\cal W}_{n n^\prime} n_{n^\prime}(t)
      -{\cal W}_{n^\prime n}
     n_n(t) \right]\nonumber\\
     -&\sum_k {\cal W}_{k n} n_n(t) +\sum_k \tau_t {\cal W}_{nk} j_k\nonumber\\
     =&\sum_m T_{nm} n_m(t) + \sum_k \tau_t {\cal W}_{nk} j_k~, 
\label{Master}
\end{align}
where $j_k\sim ke^{-\beta_eE_k}$ is the stationary flux corresponding to a single electron 
whose energy is distributed over the continuum of unbound surface states $k$ with a mean electron 
energy $k_BT_e$, ${\cal W}_{q,q'}$ is the probability per unit time for a transition from state $q'$ 
to state $q$, which can be either bound ($q',q=n$) or unbound ($q',q=k$), arising from 
the interaction with a transverse acoustic phonon, which leads to an oscillation of the image
plane, and $\tau_t=2L/v_z$ is the traveling time through the surface potential of width $L$ 
which, in the limit $L\rightarrow\infty$, can be absorbed into the transition probability per 
unit time from the continuum state $k$ to the bound state $n$, ${\cal W}_{nk}$. In 
Ref.~\cite{HBF10a} we calculated the transition probabilities per unit time ${\cal W}_{q,q'}$ 
up to fourth order in the electron-phonon coupling for a recoil corrected image potential which 
avoids the unphysical singularity of the classical image potential at $z=0$. 

The eigenvalues of the matrix ${\bf T}$ defined in Eq.~(\ref{Master}) determine the time evolution 
of the occupancies $n_n(t)$. It turns out that $n_n(t)$ contains a quickly and a slowly varying part. 
Summing the slowly varying part, which we denoted by $n^s_n(t)$, over $n$ gives the overall probability 
$n^s(t)$ of the electron to remain in any of the bound surface states after the fast energy relaxation 
within the manifold of bound and unbound surface states deceased. The overall probability satisfies 
a first order differential equation~\cite{HBF10b}, 
\begin{align}
\frac{d}{dt}n^{\rm s}(t)=\sum_k s^{\rm kinetic}_{e,k}j_k-
\frac{1}{\tau_e}n^{\rm s}(t)~,
\label{cascade}
\end{align}
with 
\begin{align}
s_{e,k}^\text{kinetic}=\tau_t \sum_{n,l} e_{n}^{(0)}
\tilde{e}_l^{(0)} {\cal W}_{lk}~,
\end{align}
the kinetic energy resolved sticking coefficient and 
\begin{align}
\tau_e^{-1}=\lambda_0
\end{align}
the electron desorption time, where $e_{n}^{(0)}$ and $\tilde{e}_n^{(0)}$ are, respectively, the 
$n^{\rm th}$ component of the right and left eigenvector corresponding to the lowest eigenvalue
$\lambda_0$ of the matrix ${\bf T}$.

Equation~(\ref{cascade}) takes cascades between bound image states and re-emission after initial 
trapping into account. Initial trapping is the transition from a continuum state $k$ to any 
bound state $n$. Its probability is given by the prompt energy resolved sticking coefficient, 
\begin{align}
s_{e,k}^\text{prompt}=\tau_t \sum_{n}
{\cal W}_{nk}~.
\end{align}

For the situation we consider, a stationary incident unit electron flux corresponding to an 
electron with Boltzmann distributed kinetic energies, it is more appropriate to discuss 
energy averaged sticking coefficients,
\begin{align}
s^{\rm ...}_e=\frac{\sum_k s^{\rm ...}_{e,k} k
e^{-\beta_e E_k}}{\sum_k k e^{-\beta_e E_k}} \text{ ,} 
\label{promptenergyavergaed}
\end{align}
where \(\beta_e^{-1}=k_BT_e\) is the mean electron energy.
\begin{figure}[t]
\centering
\includegraphics[width=1.0\linewidth]{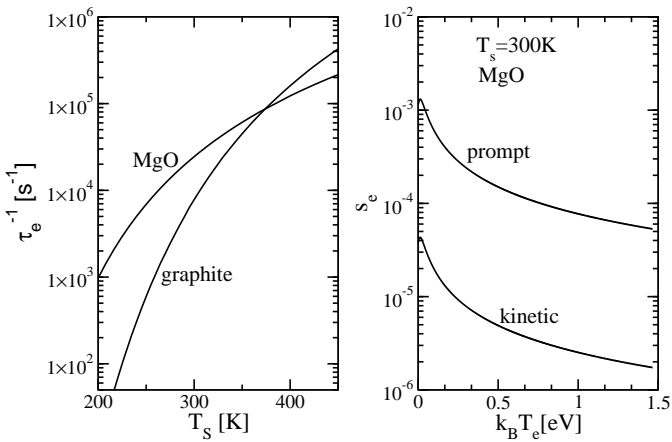}
\caption{Inverse electron desorption time for an electron thermalized in the bound
surface states of a MgO and a graphite surface (left panel) and prompt and kinetic electron 
sticking coefficient (right panel) for an electron whose kinetic energy is Boltzmann distributed
over the unbound surface states of a MgO surface.}
\label{ImagePot}
\end{figure}

Using the approach just outlined we investigated in great detail trapping~\cite{HBF10b} 
and de-trapping~\cite{HBF10a} of an electron to-and-fro an uncharged dielectric surface. Of 
particular importance is thereby the depth of the surface potential which we classified as
one-, two, and multi-phonon deep depending on whether the transition between the lowest two
bound surface states requires one-, two-, or multi-phonons. Beyond the two-phonon processes 
the calculation of the transition probabilities is very tedious. We restricted ourselves
therefore to one- and two-phonon deep potentials as it is applicable, for 
instance, to graphite and MgO. Sapphire (${\rm Al_2O_3}$), the dielectric we will consider 
in the next section, has a three-phonon deep surface potential.

In Fig.~\ref{ImagePot} we plot representative results for graphite and MgO. The 
electron desorption times vary strongly with the surface temperature $T_s$. The 
temperature dependence is exponential and can be fitted by an Arrhenius-like 
expression, $\tau_e^{-1}=cT_se^{-E_d/k_BT_s}$, with $c$ and $E_d$ fit parameters.  
The parameter $E_d$ can be interpreted as the desorption energy but it 
does not coincide with the binding energy of the lowest bound surface state as one might 
expect. The pre-exponential factor $cT_s$ is also not the frequency at the bottom of 
a potential well as it is sometimes erroneously assumed. The electron sticking coefficients
shown in the right panel of Fig.~\ref{ImagePot} are rather small, in particular, the 
kinetic sticking coefficients, which are always smaller than the prompt sticking
coefficients because they account for the possibility that the electron may de-trap 
after initial trapping. 

Empirical fits to $s_e$ and $\tau_e$ obtained from applications of the surface plasma 
model to dc column plasmas~\cite{BBD97,USB00} suggest $s_e\tau_e\approx 10^{-6}s$ 
whereas our microscopic calculation for an uncharged dielectric surface leads to 
$s_e\tau_e\approx 10^{-8}s$ or even smaller depending on the electron temperature.
The reason for the discrepancy is the neglect of the wall potential. The approach
discussed in this section is only applicable to an uncharged dielectric surface, 
where the vacuum potential is below the conduction band edge. As we have seen 
in the previous section, for a charged plasma wall, electrons already trapped on 
or in the wall lead to a Coulomb barrier for the approaching electron. 
Charge collection takes then place in an energy region where empty 
conduction band states are available. In the next section we shall discuss the
dramatic change in the physisorption microphysics which originates from this fact.

\section{Charged dielectric walls}

Plasma walls carry a negative potential of typically a few electron volts. Only
electrons with a kinetic energy large enough to overcome the Coulomb barrier 
due to this potential have a chance to come close enough to the surface to experience 
the polarization-induced attraction. In this energy range, however, polarization-induced 
image states are unstable because of the existence of empty conduction band 
states. In our notation, trapping and de-trapping of an electron no longer involves 
transitions between bound and unbound surface states but transitions between
free states and volume states. The \textquotedblleft surface
charge\textquotedblright, is thus not localized in front of the wall but occupies 
part of the interior of the wall.

The build-up of the wall potential can be still understood as an electron
physisorption process involving now, however, free states and the continuum of volume states 
in the conduction band and not the continuous and discrete spectrum of unbound and bound 
surface states. Instead of a dynamic perturbation of the surface potential, triggered 
by an acoustic phonon leading to an oscillation of the image plane, electron energy 
relaxation is now due to inelastic scattering processes within the wall, involving acoustic 
and optical bulk phonons and, if the electron energy is larger than the energy gap of the 
dielectric, impact ionization of valence electrons. Elastic scattering on impurities may also 
contribute to temporary charge trapping. 

In the following we give a rough estimate of the electron desorption time and the
electron sticking coefficient for a charged dielectric wall. A more accurate
calculation, taking a realistic electronic structure of the wall and all relevant 
scattering processes into account, will be presented elsewhere~\cite{HBF11}.
\begin{figure}[t]
\centering
\includegraphics[width=0.8\linewidth]{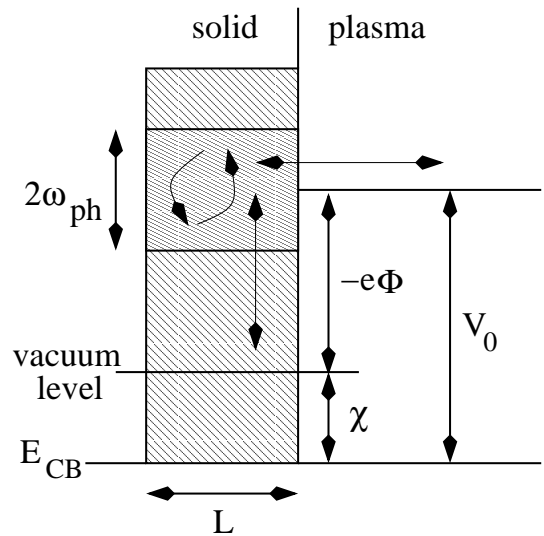}
\caption{Illustration of the model used to estimate the
electron desorption time and electron sticking coefficient for
a charged dielectric surface.}
\label{Model}
\end{figure}

The model on which our estimate is based is shown in Fig.~\ref{Model}. Motivated 
by Fig.~\ref{Sheath} we approximate the potential energy in the vicinity of the 
wall by a potential step of height $V_0=-e\phi+\chi$, where $\chi$ is the electron
affinity of the wall and $-e\phi$ is the potential energy at the wall (the 
selfconsistent wall potential (see below) we call $\phi_w$). Neglecting
multi-phonon processes, the energy interval $[V_0-\hbar\omega_{\rm ph},
V_0+\hbar\omega_{\rm ph}]$ with $\hbar\omega_{\rm ph}$ the energy of an optical
phonon is essential for trapping and de-trapping of an electron. 

Using the notation introduced in Ref.~\cite{BDF09} and briefly mentioned in the 
previous section, an electron in a free state with energy between $V_0+\hbar\omega_{\rm ph}$ 
and $V_0$ has a chance to end up via one-phonon emission in a volume state below $V_0$
(see Fig.~\ref{Model}). 
Since the thermalization in the conduction band of an insulator is extremely fast, occurring 
on the $fs$ time scale~\cite{Fitting10}, once the electron is in a state below $V_0$ it quickly  
relaxes to the lowest available volume state, which, leaving defect states aside, will 
be close to the bottom of the conduction band. Desorption from such a deep state, the 
\textquotedblleft electron binding energy\textquotedblright~is of the order of $V_0$, is 
quite unlikely. Within the one-phonon approximation, desorption can only occur if the 
trapped electron has an energy between $V_0-\hbar\omega_{\rm ph}$ and $V_0$ and absorbs 
a phonon. The probability for occupying such high-lying volume states is at room temperature
extremely small. Even without calculation we can already note that a charged 
wall will have a much longer electron desorption time than an uncharged one.

To complete the mathematical formulation of the model we need a length scale $L$ 
on which energy relaxation takes place. For the uncharged wall this was the range of the 
surface potential. Since for a charged wall energy relaxation takes place inside the 
wall this length is no longer applicable. Instead it is the penetration depth of the
electron which now determines the efficiency with which it looses energy and gets pushed 
into bound states below $V_0$. In principle, the penetration depth can be calculated from 
a Boltzmann equation taking all relevant scattering processes into account but it is
quite expensive. For the purpose of this paper, which is to discuss possible microscopic 
scenarios, we postpone such a calculation. Using the penetration depth as an adjustable 
parameter taken from experiments we can nevertheless produce reasonable first estimates 
for $s_e$ and $\tau_e$.

Ignoring cascades within the continua of free and volume states, respectively, 
the electron desorption time and the electron sticking coefficient can be obtained from 
second order perturbation theory with respect to the bulk electron-phonon coupling. More 
specifically, 
\begin{align}
\tau_e^{-1}=\langle \Gamma_{\vec{Q}q}\rangle_D
~~~{\rm and}~~~ 
s_e=\langle S_{\vec{K}k}\rangle_P
\end{align}
with
\begin{align}
\Gamma_{\vec{Q}q}=&
\sum_{\vec{K}k} {\cal W}^-(\vec{K}k,\vec{Q}q)~,\label{taueC}\\
S_{\vec{K}k}=&
\frac{2Lm^*}{\hbar k}\sum_{\vec{Q}q} {\cal W}^+(\vec{Q}q,\vec{K}k)~,
\label{seC}
\end{align}
and
\begin{align}
{\cal W}^\pm(\vec{k},\vec{k}')=&\frac{2\pi}{\hbar}\big|M(|\vec{k}-\vec{k}'|)\big|^2
\delta(E_{\vec{k}'}-E_{\vec{k}}\mp\hbar\omega_{\rm ph})\nonumber\\
\times&\big[n_B(\hbar\omega_{\rm ph})+\frac{1}{2}\pm\frac{1}{2}\big]~,
\end{align}
where 
\begin{align}
n_B(\hbar\omega_{\rm ph})=\frac{1}{\exp(\hbar\omega_{\rm ph}/k_BT_l)-1}
\end{align}
with $T_l$ the lattice temperature of the wall. The function
\begin{align}
M(|\vec{k}-\vec{k}'|)=-2i\sqrt{\frac{\hbar\omega_{\rm ph}(\epsilon_\infty^{-1}-\epsilon_s^{-1})}{2Ve^2}}
\frac{e^2}{|\vec{k}-\vec{k}'|}~.
\end{align}
is the matrix element for the scattering of a conduction band electron off a polar phonon
for vanishing conduction band electron density~\cite{Ridley99};  
$V$ is the volume occupied by the wall and $\epsilon_\infty$ and $\epsilon_s$ are,
respectively, the high frequency and the static limit of the dielectric function of 
the wall.

The energy of a bound electron, that is, an electron in a volume state, is 
$E_{\vec{Q}q}=\hbar^2Q^2/2m^*+E_q$ with $E_q=\hbar^2q^2/2m^*<V_0$, where $m^*$ is the 
effective electron mass of the conduction band and $\vec{Q}$ and $q$ are, respectively, the 
two-dimensional momentum lateral and normal to the wall. For an unbound electron, that 
is, an electron in a free state, the energy is $E_{\vec{K}k}=\hbar^2K^2/2m^*+E_k$ with 
$E_k=\hbar^2k^2/2m^*>V_0$ and $\vec{K}$ and $k$ having the same meaning for an unbound
electron as $\vec{Q}$ and $q$ for a bound one. If the unbound electron is in the
plasma halfspace the effective mass has to be replaced by the bare electron mass $m_e$.

The brackets in Eqs.~(\ref{taueC}) and (\ref{seC}) indicate averages with respect to the
weight functions $D$ and $P$, respectively. The former can be interpreted as the probability 
for a trapped electron to have the energy $E_{\vec{Q}q}$. It is given by
\begin{align}
D_{\vec{Q}q}=
\frac{\exp[-\beta_{\rm eff}E_{\vec{Q}q}]}
{\sum_{\vec{Q}'q'}\exp[-\beta_{\rm eff}E_{\vec{Q}'q'}]}
\end{align}
with an effective electron temperature $T_{\rm eff}=1/k_B\beta_{\rm eff}$. Since we 
expect electrons in the conduction band of the wall to be thermalized the effective temperature 
is equal to the lattice temperature which in turn is of the order of the room temperature and 
thus very low compared to the electron temperature $T_e$ and the potential height $V_0$. The 
weight function used in the definition of the sticking coefficient is 
\begin{align}
P_{\vec{K}k}=&\frac{\exp[-\beta_eE_{\vec{K}k}] k}
{\sum_{\vec{K}'k'}\exp[-\beta_eE_{\vec{K}'k'}] k'}~.
\end{align}

In the narrow energy range around $V_0$ where trapping and de-trapping occurs (see Fig.~\ref{Model}) 
the momentum dependence of $\Gamma_{\vec{Q}q}$ and $S_{\vec{K}k}$ is weak. We calculate therefore
both quantities only for vanishing lateral momentum (implying normal incident) and normal 
momentum equal to $(2m^*V_0/\hbar^2)^{1/2}$. Utilizing moreover that $\hbar\omega_{\rm ph}\ll V_0$ 
the integrals defining $\tau_e$ and $s_e$ can be done analytically.

Measuring energies in units of the Rydberg energy $Ry$ and lengths in units of the Bohr radius
$a_B$ and introducing a dimensionless electron-phonon coupling constant,
\begin{align}
C=&4\omega_{\rm ph}\big(\frac{1}{\epsilon_\infty}-\frac{1}{\epsilon_s}\big)~,
\end{align}
we find for the inverse electron desorption time
\begin{align}
\tau_e^{-1}=\sqrt{\frac{m^*}{m_e}\frac{\beta_{\rm eff}\omega_{\rm ph}^2}{\pi V_0^2}}
\frac{C}{8\pi}\ln\big[\frac{4V_0}{\omega_{\rm ph}}\big]\exp[-\beta_{\rm eff} V_0]\frac{Ry}{\hbar}
\label{taueCf}
\end{align}
and for the prompt electron sticking coefficient
\begin{align}
s_e=\frac{m^*}{m_e}\frac{\beta_e\omega_{\rm ph}}{V_0}
\frac{C}{8\pi}\ln\big[8\big(\frac{V_0}{\omega_{\rm ph}}\big)^2\big]\exp[-\beta_e \omega_{\rm ph}]\frac{L}{a_B}~.
\label{seCf}
\end{align}
Since we do not allow for the possibility that an initially trapped electron may desorb before it 
relaxes to the deep conduction band states we cannot distinguish between prompt and kinetic sticking.

In order to see what electron desorption times and electron sticking coefficients can be 
expected for a charged wall, we present data for a charged sapphire surface (${\rm Al_2O_3}$). 
The material parameters for sapphire are well known because of its importance for microelectronics. 
In sapphire there are two optical phonon modes which couple strongly to electrons, a longitudinal 
and a transverse one. The energy of both modes is approximately $\hbar\omega_{\rm ph}=
0.1eV$~\cite{SWK03} and the dielectric constants determining the coupling strength are for both 
modes approximately $\epsilon_\infty=3$ and $\epsilon_s=9$~\cite{STH00}. To account for the two 
modes we can thus simply multiply the transition rates by a factor two and use the given
parameter set. The effective mass of conduction band electrons in
sapphire is $m^*=0.3m_e$~\cite{SWK03}. As far as the penetration depth of electrons is concerned
we first note that after overcoming the Coulomb barrier the electrons in question have a kinetic 
energy of only a few electron volts. Measurements on ${\rm Al_2O_3}$ tunneling diodes have shown that
in this energy range electrons have penetration depths between $50\AA$ and 
$200\AA$~\cite{Pong69,Hickmott65}.
\begin{figure}[t]
\centering
\includegraphics[width=1.0\linewidth]{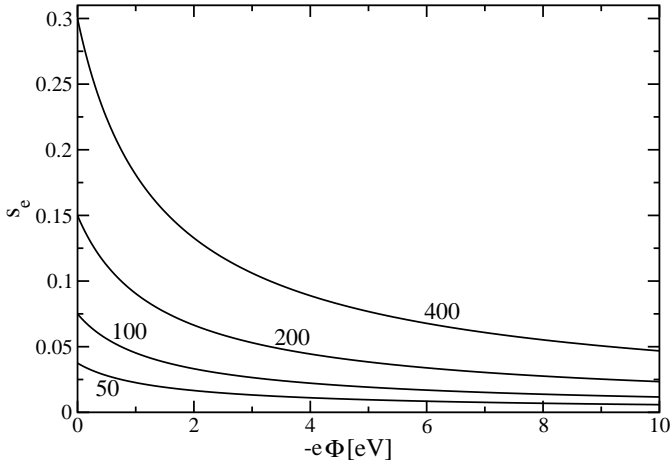}
\caption{Prompt electron sticking coefficient for a sapphire surface (${\rm Al_2O_3}$) at
room temperature as a function of the wall potential $\phi$ and the electron penetration
length $L$. The numbers attached to each graph indicate $L$ in units of the Bohr
radius $a_B$. Note, the applicability of the model on which the calculation of $s_e$ 
is based requires $-e\phi+\chi>\hbar\omega_{\rm ph}\approx 0.1 eV$ }
\label{StickCoeff}
\end{figure}

First, we discuss the electron desorption time $\tau_e$. 
As already mentioned electrons in the conduction band of an insulator thermalize with the 
lattice on a $fs$ time scale~\cite{Fitting10}. The mean energy of a trapped electron is 
thus $k_BT_{\rm eff}=k_BT_l\approx 0.026 eV$. Since, on the other hand, 
$V_0=-e\phi+\chi$ is typically a few electron volts, the exponential factor in Eq.~(\ref{taueCf})
is extremely small implying $\tau_e^{-1}\approx 0$ as assumed in the perfect absorber model.
In the initial stages of charge accumulation, however, when the wall potential  
is not yet fully developed, desorption cannot be neglected.

Let us now turn to the electron sticking coefficient. Figure~\ref{StickCoeff} shows for 
various penetration depths the electron sticking coefficient as a function of the wall 
potential. For our model to be applicable, the Coulomb barrier of the wall has to be 
larger than the phonon energy. Hence, the data shown in Fig.~\ref{StickCoeff} apply 
only to situations where $-e\phi+\chi>\hbar\omega_{\rm ph}\approx 0.1 eV$. Compared to the sticking 
coefficients of an uncharged
dielectric surface with negative electron affinity the sticking coefficients are three orders of
magnitude larger. Because energy relaxation takes now place inside the wall an initially unbound
electron couples strongly to bulk phonon modes. It can thus loose energy very efficiently 
leading to electron sticking coefficients of the order of $10^{-1}$ and not of the order of
$10^{-4}$. Note, however, in reality the sticking coefficient might be somewhat smaller
because we neglected re-emission of the electron before thermalization in the conduction 
band is completed and implicitly assumed that the transmission probability of a plasma electron 
to the solid is one whereas in reality it is energy dependent and always less than one because 
of the difference in the mass. 

Equations (\ref{taueCf}) and (\ref{seCf}) give, respectively, the electron desorption time and 
electron sticking coefficient as a function of $V_0$ and hence of $\phi$. We can thus use 
these two equations to determine the selfconsistent wall potential $\phi_w$ for a collisionless 
sheath taking surface effects beyond the perfect absorber approximation for electrons
into account. Setting $s_i=1$, $\tau_i^{-1}=\tau_e^{-1}=0$ and inserting Eq.~(\ref{seCf})
into Eq.~(\ref{balance}) gives a transcendental equation for $-e\phi$ whose root is 
$-e\phi_w$. Recall, we considered only scattering on two optical phonon modes. In reality there 
is also scattering on acoustic phonons as well as impurities which can also push electrons into 
states which are temporarily bound with respect to their normal motion. The wall potential we 
obtain is thus a lower bound to the true wall potential whereas the wall potential of the perfect 
absorber is certainly an upper bound.
\begin{figure}[t]
\centering
\includegraphics[width=1.0\linewidth]{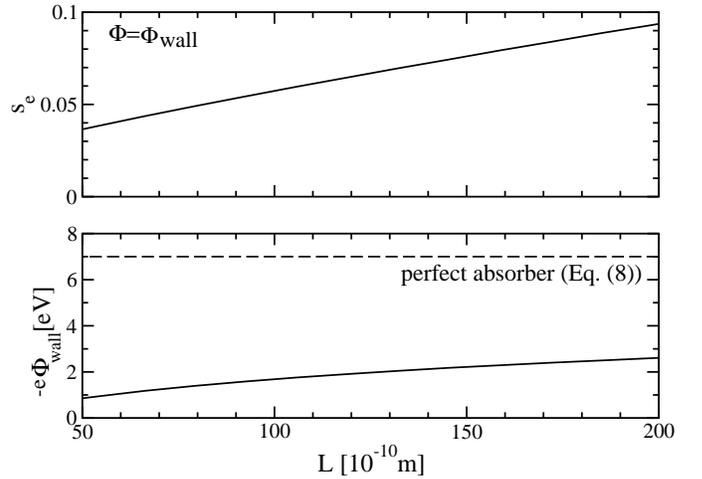}
\caption{Lower panel: Selfconsistent wall potential $\phi_w$ for a sapphire surface
in contact with a helium discharge as a function of the electron penetration length $L$.
Upper panel: Prompt electron sticking coefficient at the selfconsistent wall potential.
The mean electron energy in the discharge is $k_BT_e=2eV$.}
\label{PotWall}
\end{figure}

As can be seen in the lower panel of Fig.~\ref{PotWall}, the wall potential including 
surface effects for electrons is roughly one-third of the wall potential of the perfect 
absorber, Eq.~(\ref{phiw}). The true wall potential should be somewhere between our result 
and the perfect absorber value. The accuracy of our theoretical estimate is of course 
not good enough to make more precise statements. The same may be unfortunately said about 
experimental measurements. Nevertheless it is encouraging that the 
approximate expressions (\ref{taueCf}) and (\ref{seCf}) produce in conjunction 
with (\ref{balance}) wall potentials of the expected order of magnitude. The upper
panel of Fig.~\ref{PotWall} finally shows that the electron sticking coefficient of a charged 
wall is of the order of $10^{-1}$ and thus significantly smaller than assumed in the perfect
absorber model.

\section{Conclusion}

The purpose of this paper was to discuss the interaction of plasma electrons 
with plasma walls beyond the perfect absorber approximation. Instead of assuming 
an electron hitting the wall to be absorbed with certainty and never released 
again we proposed a physisorption-inspired quantum-mechanical model to calculate 
the probability with which an electron gets stuck to the plasma wall -- the 
electron sticking coefficient $s_e$ -- and the time the electron remains on or 
in the wall -- the electron desorption time $\tau_e$. 

The microphysics controlling $s_e$ and $\tau_e$ depends on the charge of the wall.
When the wall is uncharged, that is, in the early stages of the charging process
and has a negative electron affinity
sticking and desorption occurs in polarization-induced external bound surface 
states (image states) and is triggered by inelastic scattering cascades with acoustic
phonons. The sticking coefficient $s_e$ is then very small, at most of the order of $10^{-4}$, 
and the desorption time $\tau_e\approx 10^{-4}s$. The wall is thus far from being
a perfect absorber for electrons which would correspond to $s_e=1$ and $\tau_e^{-1}=0$. 

Once the wall is charged, the negative wall potential $\phi$ blocks surface and 
volume states between the vacuum level and the Coulomb barrier. An approaching electron 
overcoming the Coulomb barrier may then directly enter empty conduction band 
states, that is, volume states which do not exponentially decay inside the bulk of 
the wall as image states do. Electron energy relaxation due to inelastic scattering 
with optical bulk phonons may then be very efficient in pushing the electron below 
the Coulomb barrier. As a result, it gets stuck. Once it is stuck, thermalization 
with the lattice is very fast implying that the stuck electron relaxes quickly to the 
bottom of the conduction band from which it cannot escape at room temperature. 
Within this scenario the binding energy of the trapped electrons is approximately 
$-e\phi+\chi$, where $\phi$ is the actual wall potential and $\chi$ is the electron affinity 
of the wall, $s_e$ is of the order of $10^{-1}$ and $\tau_e^{-1}\approx 0$. Hence, if it was 
not for $s_e$, the wall would be a perfect absorber.

Calculating $s_e$ and $\tau_e$ as a function of the wall potential $\phi$ and inserting 
these two expressions in the quasi-stationary balance equations for the electron and 
ion surface densities of a collisionless sheath while assuming the wall to be a perfect
absorber for ions we obtained the selfconsistent wall potential $\phi_w$ beyond the 
perfect absorber approximation for electrons. Taking electron surface effects into account 
reduces $\phi_w$ approximately by a factor three compared to the perfect 
absorber value. 

Our investigation clarifies the materials science aspects which have to be resolved in order 
to go beyond the perfect absorber model for electrons. The most important one is of course 
the precise electronic structure of the wall including defect states due to surface 
reconstruction and/or chemical contamination because it determines the nature of the 
states which potentially host the electrons building up the wall
potential. But also the thermalization and penetration
of electrons with only a few electron volts kinetic energy are critical processes. From 
low energy electron diffraction it is known that in this energy range the interaction of 
electrons with solids is particularly intricate. Although the accuracy of the perfect absorber 
model for electrons might be sufficient for the modeling of traditional electrical discharges, 
for the modeling of dusty plasmas and solid-state-based microdischarges the description 
of the electron-wall interaction along the lines presented here will be vital.


%



\section*{Acknowledgment}

Support from the Deutsche Forschungsgemeinschaft through the 
transregional collaborative research center TRR 24 is greatly
acknowledged. J.M. is supported by the International 
Max-Planck Research School for bounded plasmas.

\ifCLASSOPTIONcaptionsoff
  \newpage
\fi



\bibliographystyle{IEEEtran}
\end{document}